# Numerical Simulation of Asymmetric Merging Flow in a Rectangular Channel

Abuzar A. Siddiqui

*Department of Basic Sciences, Bahauddin Zakariya University, Multan, Pakistan*

## Abstract

*The steady, asymmetric and two-dimensional flow of viscous, incompressible, and Newtonian fluid through a rectangular channel with splitter plate parallel to walls is investigated numerically. In the past, the position of the splitter plate was taken as a centerline of channel but here we consider its different positions which cause the asymmetric behaviour of the flow field. The geometric parameter that controls the position of splitter is defined as splitter position parameter $\alpha$.. The plane Poiseuille flow is considered far from upstream and downstream of the splitter. The fourth order method Mancera [11], followed by special finite-method Dennis [5] is used to solve in our purpose. This numerical scheme transforms the governing equations to system of finite-difference equations which are solved by point S.O.R. iterative method. Moreover, the results obtained are further refined and upgraded by Richardson Extrapolation method Jain [18]. This scheme yields the sixth order accurate solution. The calculations are carried out for the ranges $-1<\alpha<1$, and $0 \leq R < 10^5$. The results are compared with those of Badr et al [1] for symmetric case (when $\alpha=0$) for velocity, vorticity and skin friction distributions. The comparison is very favourable.*

## 1. Introduction

The mixing of two different fluids flows and concentration at a junction of two channels constitutes the field of concentration. It is also called merging flow. It has great interest in physical application point of view. For example Sayre [14] in his consideration of problem of the mixing of two rivers assumed that the concentration field was completely unmixed at the point of confluence. It has a lot of use in physiological flows, internal machinery dynamics, lakes, estuaries, and rivers, see for example: Badr et al [1], and Hamblin [6]. This type of geometry is used in quadrupole magnetic cell sorter William[17] and is also dominant in respiratory flow Sera [15], and Pedley [13].

In this paper, we consider the steady, two-dimensional flow of incompressible viscous fluid that constitutes the merging flow in which two oncoming channel flows are forced to join into one. The geometry of problem consists of an infinite long straight channel $|y'| \leq h$, $-\infty < x' < \infty$ of width $2h$ containing a disposed straight splitter plate placed at $y'=\alpha h$, $-1<\alpha<1$ within the channel as shown in figure 1. Here $x'$, $y'$ represent dimensional rectangular coordinates. The position of splitter is controlled by a splitter position parameter, $\alpha$. If $\alpha=0$ then the splitter plate will coincide with the x-axis while the plate will lie above and below the x-axis if $\alpha>0$ and $\alpha<0$ respectively. The inlet and outlet boundary flow assumed to be same as the plane Poiseuille flow appropriate to the single width channel for downstream while double channel for upstream of splitter. Although for symmetric case, the various attempts have been made by several authors for example; Badr [1], Bramley [2-4], Krijger [9], Lonsdale [10], Nakamura [12], Tsui [16] yet owing to the complexity for asymmetric flow which resembles actually to the real life problems unfortunately due attention is not given.

Obviously a question arises how does differ our study with previous attempts? Following are few facts which cover the answer of this question.

- We consider non-uniform rectangular meshing while they used mostly square grid.
- We use higher order scheme while previously the authors used mostly $2^{nd}$ order scheme.
- We consider asymmetric flow while previously the symmetric flow cases were examined.
- We study the low, moderate, and high Reynolds numbers flow while previous attempts are mostly based either on low, moderate or high Reynolds number see for example Badr et al [1],
- Further in the paper by Badr [5] a little correction is made which is mentioned in section 5.

The main aim of this attempt is to explore all the aspects of the flow behaviour and examine the understanding of the complicated features such as boundary layer separation, growth and decay of vortices, reattachment, trailing-edge and leading-edge properties three dimensionally and variation of skin friction on the sold walls and splitter plate for asymmetric case. In addition to this, the flow behaviour is also simulated for various values of Reynolds number R and splitter position parameter, $\alpha$. We also observe how asymmetric study differs from the previously attempted symmetric cases. We examine the variation of eddies when the splitter plate is moved upward or downward within the channel. The extrapolated results obtained are simulated graphically and are compared with those of previously attempted symmetric cases e.g. Badr et al [1] etc. These are given in section 5. The formulation and basic flow analysis are presented in section 2 while sections 3 and 4 contain the detail of numerical scheme so adopted and computational procedure used respectively.

## 2. *Basic Analysis*

The continuity and the Navier-Stokes equations for steady flow of incompressible fluid in dimensional form, in the absence of body force are:

$$\nabla \cdot \mathbf{V}' = 0 \qquad (2.1)$$

$$\rho(\mathbf{V}' \cdot \nabla')\mathbf{V}' = -\nabla' p' + \mu \nabla'^2 \mathbf{V}' \qquad (2.2)$$

where $\mathbf{V}'$ is the fluid velocity vector, $\rho$ being the constant density, $\mu$ the viscosity coefficient, and $p'$ the pressure.

The mechanics of the problem under consideration can briefly be stated as; flow is through an infinite length uniform channel with finite width parallel walls. The upstream region is divided into two channels with the help of semi-infinite splitter plate while downstream is flow between two plates without any splitter plate. The walls and splitter are parallel to each other and x-axis. The upstream of splitter plate is separated by $2\alpha h$ while downstream is separated by $2h$, as shown in figure 1. Moreover, the flow is being considered as steady, two-dimensional, laminar, and asymmetric about x-axis in general. The flow is generated by the uniform pressure gradient analogous to that of plane Poiseuille flow on inlet and outlet flow to preserve the continuity.

On deforming equations (2.1) – (2.2) into vorticity transport equation in dimensionless form by normalizing velocity, space coordinates and vorticity vector by U (main stream velocity), h, and h/U respectively, we will obtain,

$$\nabla \cdot \mathbf{V} = 0 \qquad (2.3)$$

$$\nabla \times \nabla \times \boldsymbol{\omega} = R[\nabla \times (\mathbf{V} \times \boldsymbol{\omega})] \qquad (2.4)$$

where $\boldsymbol{\omega}=[0, 0, E]$ is the vorticity vector, R the Reynolds number which is equal to the ratio $Uh/\nu$ and $\nu$ being the kinematic viscosity.

If we introduce, $u = \dfrac{\partial \psi}{\partial y}, v = -\dfrac{\partial \psi}{\partial x}$ as velocity components in terms of stream function $\psi$ relation into equations (2.5) – (2.6), we get,

$$\nabla^2 E = R\left[\frac{\partial \psi}{\partial y}\frac{\partial E}{\partial x} - \frac{\partial \psi}{\partial x}\frac{\partial E}{\partial y}\right] \qquad (2.5)$$

where $E = -\nabla^2 \psi$ \qquad (2.6)

For the geometry of figure 1 the boundary conditions for these equations are given as follows:
(a) No slip at the walls, i.e. u=v=0, on all the solid walls and the splitter plate. \qquad (2.7a)

(b) 
$$\begin{aligned}
&\text{(i) For } \alpha > 0, \quad \psi \to \frac{1}{(1-\alpha)^3}\left[-2y^3 + 3y^2 + 3\alpha(y^2 - 2y + \alpha) - \alpha^3\right], \\
&\qquad\qquad\qquad E \to \frac{1}{(1-\alpha)^3}[12y - 6 - 6\alpha] \\
&\text{(ii) For } \alpha < 0, \quad \psi \to \frac{-1}{(1+\alpha)^3}\left[2y^3 + 3y^2 - 3\alpha(y^2 + 2y - \alpha) + \alpha^3\right], \\
&\qquad\qquad\qquad E \to \frac{1}{(1+\alpha)^3}[12y + 6 - 6\alpha]
\end{aligned}$$
as $x \to -\infty$ \qquad (2.7b)

(c) $\psi \to 1.5y - 0.5y^3, E \to 3y$ \qquad\qquad as $x \to \infty$ \qquad (2.7c)

Here equations (2.7b) and (2.7c) are due to imposed Poiseuille flow far upstream and far downstream of the splitter. Note that if we put $\alpha=0$ in the boundary conditions given in (2.7) then it will become boundary condition of symmetric flow case Badr et al [1]. To solve the above boundary value problem, we use the numerical schemes given in section 3, and that reduces the highly non-linear system of partial differential equations to system of difference equations. This system, having the associated matrix which is diagonally dominant, is solved by the S. O. R. - iterative method that accelerates the convergence of the iterative scheme. Henceforth this numerical procedure is efficient, straightforward, economical in core storage requirements of a computer, and easy to programme.

## 3. *The Numerical Scheme*

To overcome the problem of rate of convergence and stability of the numerical scheme, we adopt the numerical scheme which consists of the following two methods.

The grid size along x and y directions, will be taken by H, $K_1$ respectively, and around a typical internal grid point $(x_0, y_0)$ we adopt the convention that quantities at $(x_0, y_0)$, $(x_0+H, y_0)$, $(x_0, y_0+K_1)$, $(x_0-H, y_0)$, $(x_0, y_0-K_1)$, $(x_0+2H, y_0)$, $(x_0, y_0+2K_1)$, $(x_0-2H, y_0)$, and $(x_0, y_0-2K_1)$ are represented by the subscripts 0, 1, 2, 3, 4, 5, 6, 7 and 8 respectively.

## (I) Fourth order Finite-Difference Method

If we employ the method similar to Mancera [11], then governing equations (2.5 – 2.6) would deform as:

$$\frac{1}{12H^2}[16\psi_1 + 16\psi_3 - \psi_5 - \psi_7] + \frac{1}{12K_1^2}[16\psi_2 + 16\psi_4 - \psi_6 - \psi_8] - \frac{5}{2}[\frac{1}{H^2} + \frac{1}{K_1^2}]\psi_0 = -E_0 \quad (3.1)$$

$$\left(\frac{4}{3H^2} + 8P_0\right)E_1 + \left(\frac{4}{3K_1^2} + 8Q_0\right)E_2 + \left(\frac{4}{3H^2} - 8P_0\right)E_3 + \left(\frac{4}{3K_1^2} - 8Q_0\right)E_4 - \left(\frac{1}{12H^2} + P_0\right)E_5$$

$$-\left(\frac{1}{12K_1^2} + Q_0\right)E_6 - \left(\frac{1}{12H^2} - P_0\right)E_7 - \left(\frac{1}{12K_1^2} - Q_0\right)E_8 - \frac{5}{2}\left(\frac{1}{H^2} + \frac{1}{K_1^2}\right)E_0 = 0 \quad (3.2)$$

where

$$P_0 = \frac{-R}{144HK_1}(-\psi_6 + 8\psi_2 - 8\psi_4 + \psi_8) \quad \text{and} \quad Q_0 = \frac{R}{144HK_1}(-\psi_5 + 8\psi_1 - 8\psi_3 + \psi_7)$$

## (II) Special Finite-Difference Method

In order to approximate linear partial differential equation (2.6), we employ second order standard central difference formulation at point "0", we obtain,

$$\frac{1}{H^2}\psi_1 + \frac{1}{K_1^2}\psi_2 + \frac{1}{H^2}\psi_3 + \frac{1}{K_1^2}\psi_4 - [\frac{2}{H^2} + \frac{2}{K_1^2}]\psi_0 = -E_0 \quad (3.3)$$

To deal with equation (2.5), we split it equation into following two equations as:

$$\frac{\partial^2 E}{\partial x^2} + B\frac{\partial E}{\partial x} = A(x, y) \quad (3.4)$$

$$\frac{\partial^2 E}{\partial y^2} + C\frac{\partial E}{\partial y} = -A(x, y) \quad (3.5)$$

where **A** is an arbitrary function of x and y while,

$$B = -R\frac{\partial \psi}{\partial y}, \qquad C = R\frac{\partial \psi}{\partial x} \quad (3.6)$$

Equation (3.4) is then approximated along the grid line $y=y_0$, over the range $x_0-H \leq x \leq x_0+H$, by applying a local transformation for E, given as follows:

$$E = \lambda e^{-F} \quad (3.7)$$

$$\text{where,} \quad F = \frac{1}{2}\int_{x_0}^{x} B(t, y)dt \quad (3.8)$$

On the introduction of equation (3.7), into equation (3.4) and approximation of the derivatives involved by $2^{nd}$ order central differences, we get,

$$(\lambda_1 + \lambda_3 - 2\lambda_0) - [\frac{1}{2}(\frac{\partial B}{\partial x})_0 + \frac{1}{4}B_0^2]\lambda_0 H^2 = A_0 H^2 \quad (3.9)$$

Similarly, equation (3.5) is approximated along the grid line $x=x_0$, over the range $y_0-k_1 \leq y \leq y_0+k_1$, by applying a local transformation for E, given as follows,

$$E = \mu e^{-G} \quad (3.10)$$

$$\text{where,} \quad G = \frac{1}{2}\int_{y_0}^{y} C(x, t)dt \quad (3.11)$$

On using equation (3.10) into equation (3.5), after the similar approximations as above, we shall obtain,

$$\frac{H^2}{K_1^2}(\mu_2 + \mu_4 - 2\mu_0) - [\frac{1}{2}(\frac{\partial C}{\partial y})_0 + \frac{1}{4}C_0^2]\mu_0 H^2 = -A_0 H^2 \tag{3.12}$$

On eliminating $A_0$ between equations (3.9) and (3.12), we have,

$$[\lambda_1 + \lambda_3 - 2E_0 - \frac{1}{4}E_0 B_0^2 H^2] + \frac{H^2}{K_1^2}[\mu_2 + \mu_4 - 2E_0 - \frac{1}{4}E_0 C_0^2 H^2] = 0 \tag{3.13}$$

In order to express $\lambda$ and $\mu$ back in terms of E, it can be done from the definitions. It is found that,

$$\lambda_i = E_i e^{F_i}, \qquad \mu_j = E_j e^{G_j} \tag{3.14}$$

where, $F_i = \frac{1}{2}\int_{x_0}^{x_i} B(t, y_0) dt, \quad G_j = \frac{1}{2}\int_{y_0}^{y_j} C(x_0, t) dt$, \hfill (3.15)

where i=1, 3 and j=2, 4.

Now, we expand above exponential in powers of their arguments and keep the truncation error of order $H^4$ and $H^2 K_1^2$. After some simplifications under above arguments and using the Taylor's theorem during calculations, we obtain,

$$\lambda_1 + \lambda_3 = [1 + \frac{H^2}{4}(\frac{\partial B}{\partial x})_0 + \frac{B_0^2 H^2}{8}][E_1 + E_3] + \frac{B_0 H}{2}[E_1 - E_3] \tag{3.16}$$

$$\mu_2 + \mu_4 = [1 + \frac{K_1^2}{4}(\frac{\partial C}{\partial y})_0 + \frac{C_0^2 K_1^2}{8}][E_2 + E_4] + \frac{C_0 K_1}{2}[E_2 - E_4] \tag{3.17}$$

On using equations (3.16) and (3.17) into equation (3.13), we get,

$$[1 + \frac{B_0 H}{2} + \frac{B_0^2 H^2}{8}]E_1 + [\frac{H^2}{K_1^2}\{1 + \frac{K_1^2 C_0^2}{8}\} + \frac{C_0 H^2}{2K_1}]E_2 + [1 - \frac{B_0 H}{2} + \frac{B_0^2 H^2}{8}]E_3 +$$

$$[\frac{H^2}{K_1^2}\{1 + \frac{K_1^2 C_0^2}{8}\} - \frac{C_0 H^2}{2K_1}]E_4 - [2 + \frac{2H^2}{K_1^2} + \frac{B_0^2 H^2}{4} + \frac{C_0^2 H^2}{4}]E_0 = 0, \tag{3.18}$$

These approximations give rise to an associated matrix that is always diagonally dominant. The present formulation was found to work well for all values of Reynolds number R, whereas results could only be obtained for relatively small values of R by using standard central-difference approximations to equations (2.5) and (2.6). Finally the condition for E is required at grid points on the solid walls. Here we use the same approximation as that is originally due to Woods [18] and is given by,

$$E_{w_1} = \frac{3}{K_1^2}[1 - \psi_a] - \frac{1}{2}E_a, \qquad \text{on the upper wall} \tag{3.19a}$$

$$E_{w_2} = -\frac{3}{K_1^2}[1 + \psi_b] - \frac{1}{2}E_b, \qquad \text{on the lower wall} \tag{3.19b}$$

$$E_{w_3} = -\frac{3}{K_1^2}\psi_c - \frac{1}{2}E_c, \qquad \text{on the splitter plate} \tag{3.19c}$$

where the subscripts $w_1$, $w_2$, $w_3$, represent for the value at the approximate boundary points on upper, lower, and splitter plate respectively while the subscripts a., b, and c signify to the internal grid point most immediate to $w_1$, $w_2$, $w_3$ respectively.

## *4. Computational Procedure*

This numerical experiment consists of two steps. In the first step, the Equations (3.3) and (3.18) are solved iteratively with respect to the boundary conditions (3.19) by SOR-method Hildebrand [7] until the convergence is achieved according to the criterian:

$$\max\left|E_0^{(m+1)} - E_0^{(m)}\right| < 10^{-5}, \qquad \max\left|\psi_0^{(m+1)} - \psi_0^{(m)}\right| < 10^{-5} \qquad (4.1)$$

In the second step, the equations (3.1) and (3.2) are solved numerically with respect to boundary conditions (3.19) by SOR-method on using the data obtained from step 1 as an initial estimation. This procedure is repeated until the convergence is attained according to the above criterion (4.1). The solutions obtained are further extrapolated by Richardson's extrapolation technique Jain [8].

## *5. Results, Comments and Comparisons*

Two parameters R and α represent the general fluid motion. The ranges for R and α are $0 \leq R \leq 10^5$ and $-1 < \alpha < 1$ but, for the purpose of presentation and discussion we consider R=1, 10, 100, 500, 1000, 100000 while α=-0.5, 0, 0.2, 0.5 throughout the study. The numerical calculations are carried out on grid sizes H=1/20, 1/40, 1/80, 1/160, 1/320 and $K_1$= 1/60, 1/120, 1/240, 1/480 and the presented results are extrapolated using Richardson Extrapolation to the limit. In all the cases the infinite length of channel is numerically approximated and is taken as four.

The streamlines for R=1, 10, 100, 500, 1000, 100000 and α=-0.5, 0, 0.2, 0.5 are presented in figures 2 to 6. It can be easily observed that the vortices are produced in the vicinity of the trailing edge of the splitter plate, which is singular for vorticity, (that confirms the exactness of our calculations). The size of eddies increases as α increases or if the splitter plate moves upward towards upper plate as shown in figures 2 to 6. An interesting situation exists for R=100000 and α=0.5 where whole fluid below the splitter plate is rotating clockwise as shown in fig 6(d).

The trailing edge of the splitter plate needs special attention due to the infinite or singular behaviour of vorticity at this point. To deal with this singular point two special treatments are being adopted during calculations for all values of R, and α. One is the method very similar to of Bramley [2-3] and second is that in the vicinity of the splitter plate and specially its trailing edge, the refine meshing are taken as shown in figure 13. More clustering near the singular point is needed as R increases.

Here it is necessary to comment on figure 2 of Badr [1] that our results for $u_\alpha$ (velocity at centre line) for symmetric flow match well for low Reynolds number and slightly differ for high R near downstream end of the splitter (in our case it is x=2) where the matching boundary condition $\psi \to 0.5y(3 - y^2)$ as $x \to \infty$ is applicable and son the velocity will be 1.5(1-$y^2$). For symmetric case the velocity on the central line ($u_{cl}$ in [1] and $u_\alpha$: α=0 in our case) will be obviously 1.5 for all R. This fact is missing in [1] and this deviation is corrected and is displayed in figure 7(b). In figure 7, the distribution for velocity $u_\alpha$ along the splitter plate is given for various values of Reynolds number and the splitter position parameter considered. When α=0 it is observed that the velocity along the splitter increases in downstream of trailing edge of the splitter as we move away from it and curvature of velocity profile increases as R increases. This situation will exist till R=1000 but it reverses dramatically when R>1000 as shown in figure 6(b). It is noted that the optimum value of velocity lies as x increases far from the trailing edge of the splitter, for symmetric case but it is not so for asymmetric case as shown in figures 6(a-d). Moreover, the velocity increases at once after passing the trailing edge for asymmetric case which is clear in figure 6.

The vortex effects or in other words the variation of the skin friction on the upper and lower plates of the channel are also examined here and are displayed in figures 8 and 9 for various values of R and $\alpha$ considered. One effect can be noted that the vorticity increases as R increases and it decreases as $\alpha$ increases. These effects occur on both the walls of the channel as shown in figures 8 and 9. Figure 10 (a-d) shows the variation of dimensionless vorticity on the splitter for various values of R and $\alpha$ considered. Figure 10(b) belongs to symmetric flow case and it coincides and compares well with figure 1 of Badr [1]. This verifies our results. Moreover, the main effect here is that the decay of vorticity to its downstream value takes place over an increasingly longer scale of x as R increases for symmetric case but it is not so for asymmetric one, as shown in figure 10.

The coefficient of skin friction $C_f$ has vital role in the motion of the fluids and loading effects on the solid walls. It generally depends upon vorticity, Reynolds number, viscosity, and other geometric aspects. This effect has also been studied here for various values of R and $\alpha$ considered and its variation with respect to x is also displayed graphically in figure 11. This figure indicates that the shear stress on the splitter increases as x increases and as R decreases. The maximum shear stress will occur at the singular point, which resembles with the physical reality. Figure 11 represents the comparison of our numerical solutions with analytical estimation of Badr [1]. The curves for analytical results are displayed as dotted line. The comparison is seemed to be satisfactory. Moreover, it is observed that as the splitter plate is moved upward then the shear stress on the splitter plate increases as shown in figure 11.

For comparison purpose, figure 12 indicates the variation of vorticity on the splitter for symmetric case for the various values of R considered. The curve with '*' represents the asymptotic solution of Badr [1] while other curves with solid lines belong to our numerical solutions.

## **References**

[1] Badr, H., Dennis, S.C.R., Smith, F. T. (1985). Numerical and asymptotic solutions for merging flow through a channel with an upstream splitter plate. J. Fluid Mech., vol. 156, P63-81.

[2] Bramley, J. S. (1982). A numerical treatment of two-dimensional flow in a branching channel. In Lecture notes in Physics vol. 170, P155, springer.

[3] Bramley, J. S. (1984). The numerical solution of two-dimensional flow in a branching channel. Comp. Fluids, vol. 12(b), P339-355.

[4] Bramley, J. S. (1987). Numerical solution for two-dimensional flow in a branching channel using boundary-fitted coordinates. Comp. Fluids, vol. 15(3), P297-311.

[5] Dennis, S.C.R., Smith, F. T.(1980). Steady flow through a channel with a symmetrical constriction in the form of step. Proc. R. Soc. Lond. A 372, 393.

[6] Hamblin, P. F. (1980). An analysis of advective-diffusion in branching channels. J. Fluid Mech., vol. 99, part1, P101-110.

[7]Hildebrand, F. B.(1978).Introduction to Numerical Analysis, Tata McGraw Hill Publishing Co. Ltd.

[8] Jain, M.K. (1997). Numerical Methods for Scientific and Engineering Computation. New Age International (P) Limited, Publishers.

[9] Krijger, J. K. B., Hillen, B. (1990). Steady two-dimensional merging flow from two channels into a single channel. Applied Scientific Research, vol. 47(3), P233-246.

[10] Lonsdale, G., Bramley, J. S. (1988). A non-linear multigrid algorithm and boundary-fitted coordinates for the solution of a two-dimensional flow in a branching channel. J. Comp. Phy., vol. 78(1), P1-14.

[11] Mancera, P.F., Hunt, R.(1997). Fourth order method for solving the Navier-Stokes equations in a constricting channel. Int. J. Num. Meth. Fluids., 25, 1119.

[12] Nakamura, K., Chiba, K. (1993). Numerical simulation of fibre suspension flow. Part 1. Merging flow. J. Textile Mach., Society of Japan-English edition, vol. 39(1), P1-6.

[13] Pedley, T.J. (1980). The Fluid Mechanics of Large Blood Vessel. Cambrige university press.

[14] Sayre, W. W. (1973). Natural mixing process in rivers. Environmental impact on rivers (ed. Shen Hrech Wen.) cha 6. Fort Collins, Colorado, U.S.A..

[15] Sera, T., Sanao. S, Hirohisa, H. (2005). Respiratory in a realistic tracheostenosis model. J. Biomechanical Engg., vol. 125(4), P461-471.

[16] Tsui, Y. Y.,Lu, C. Y. (2005). A study of the recirculating flow in planar, symmetrical branching channels. Int. J. Num. Meth. Fluids, vol. 50(2), P235-53.

[17] William, P. S. (1999). Flow rate optimization for the quadrupole magnetic cell sorter. Anal. Chem., vol. P3799-3807.

[18] Woods, L.C.(1954). A note on the numerical solutions of fourth order differential equations. Aeronaut. Q. 5, 176.

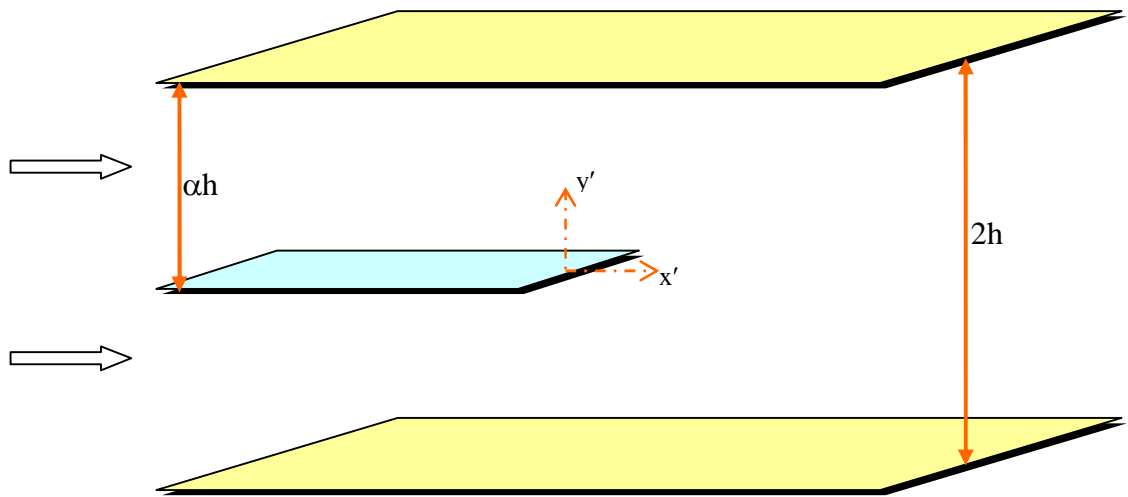

Figure 1. The Definition Sketch.

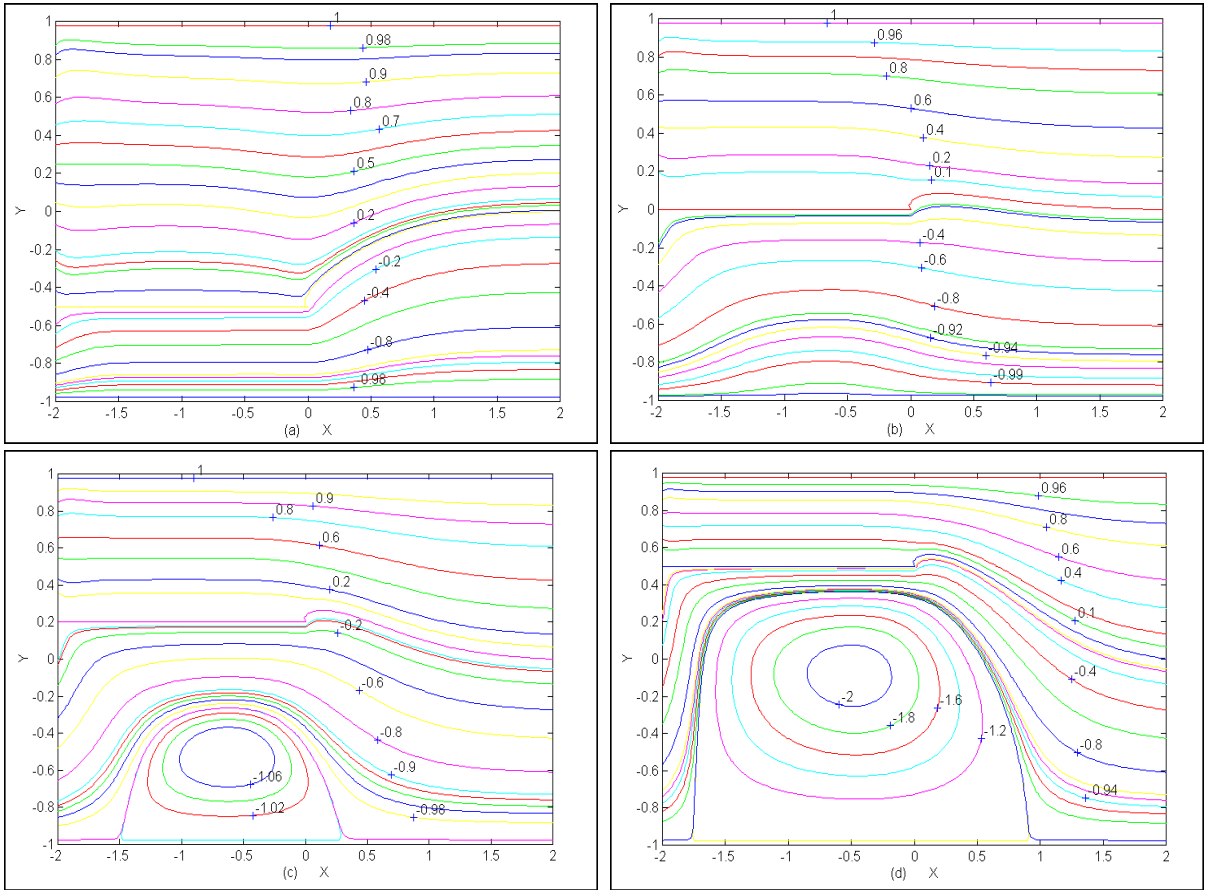

Figure 2. Streamlines for L'=4, R=1, (a) α=-0.5, (b) α=0, (c) α=0.2, (d) α=0.5.

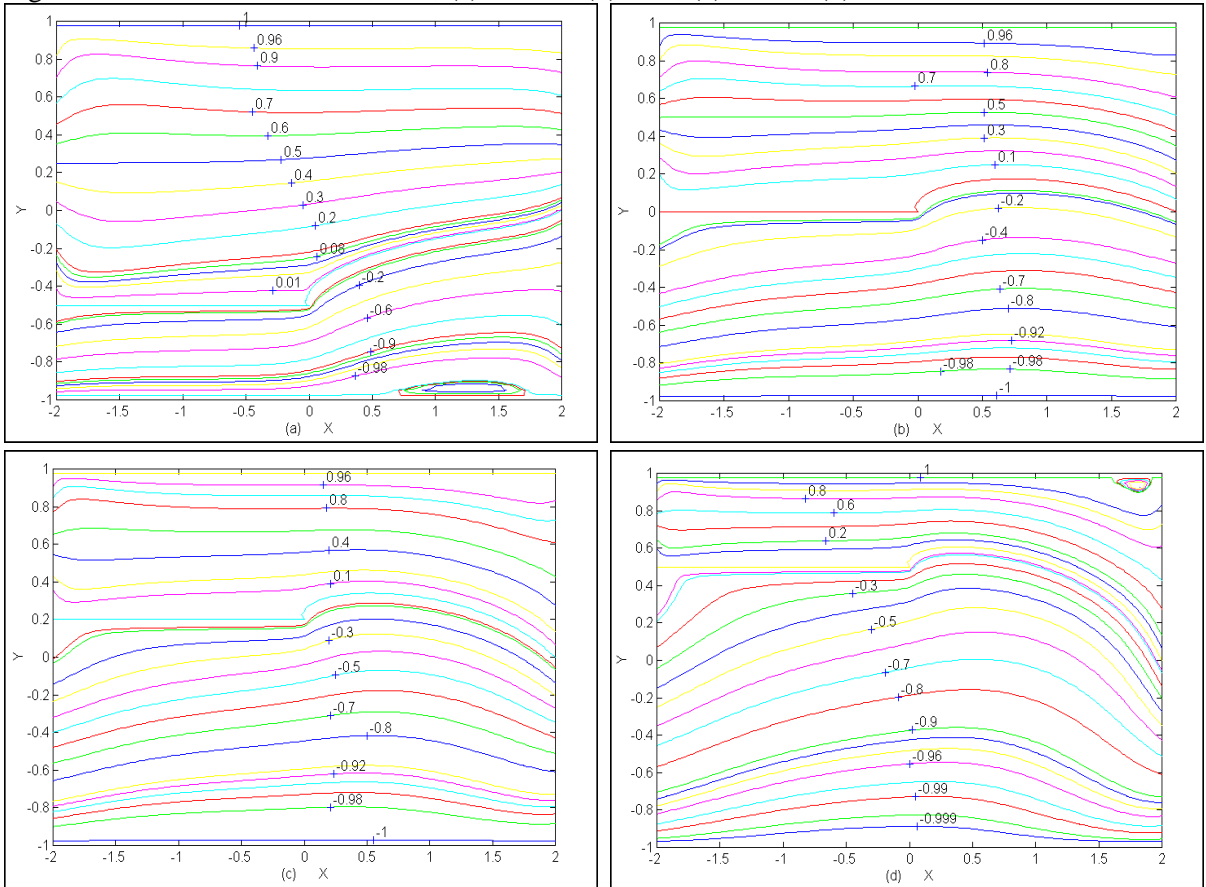

Figure 3. Streamlines for L'=4, R=100, (a) α=-0.5, (b) α=0, (c) α=0.2, (d) α=0.5.

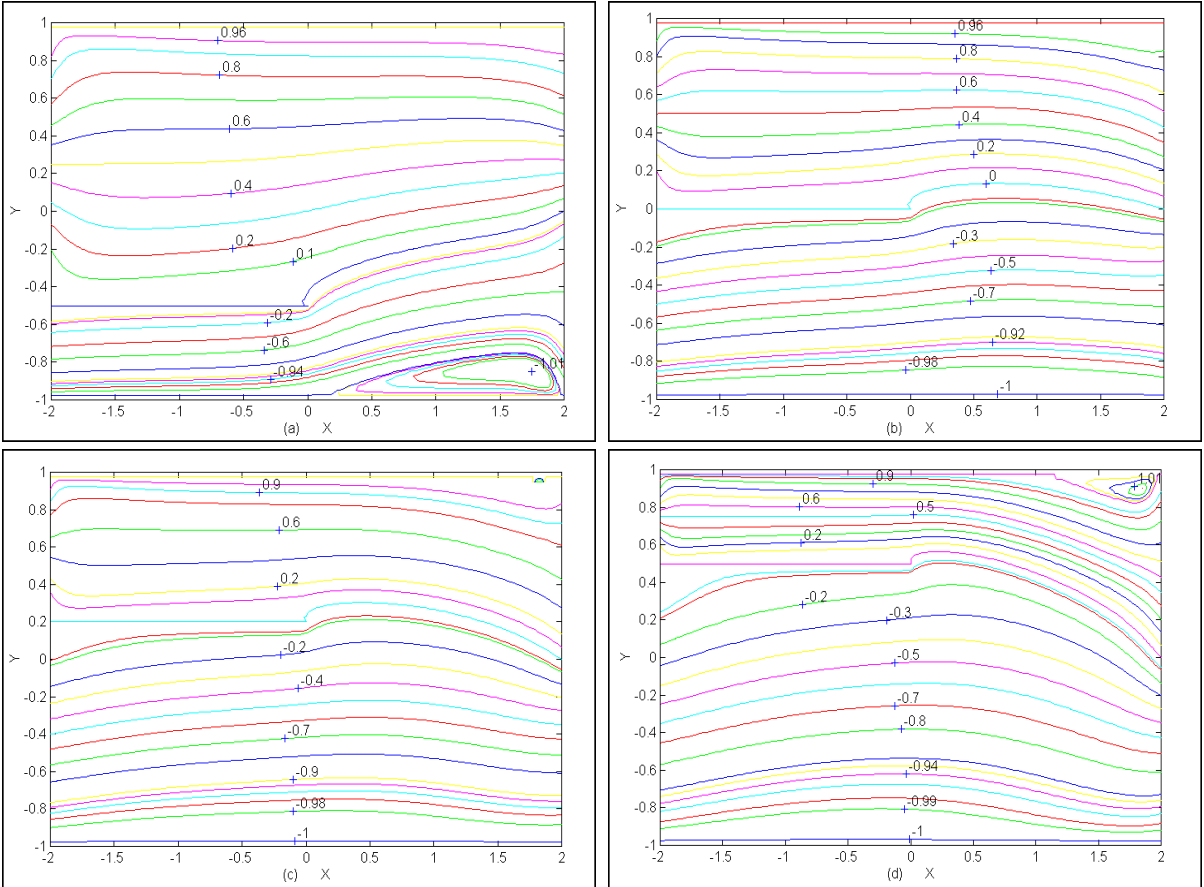

Figure 4. Streamlines for $L'=4$, R=500, (a) $\alpha=-0.5$, (b) $\alpha=0$, (c) $\alpha=0.2$, (d) $\alpha=0.5$.

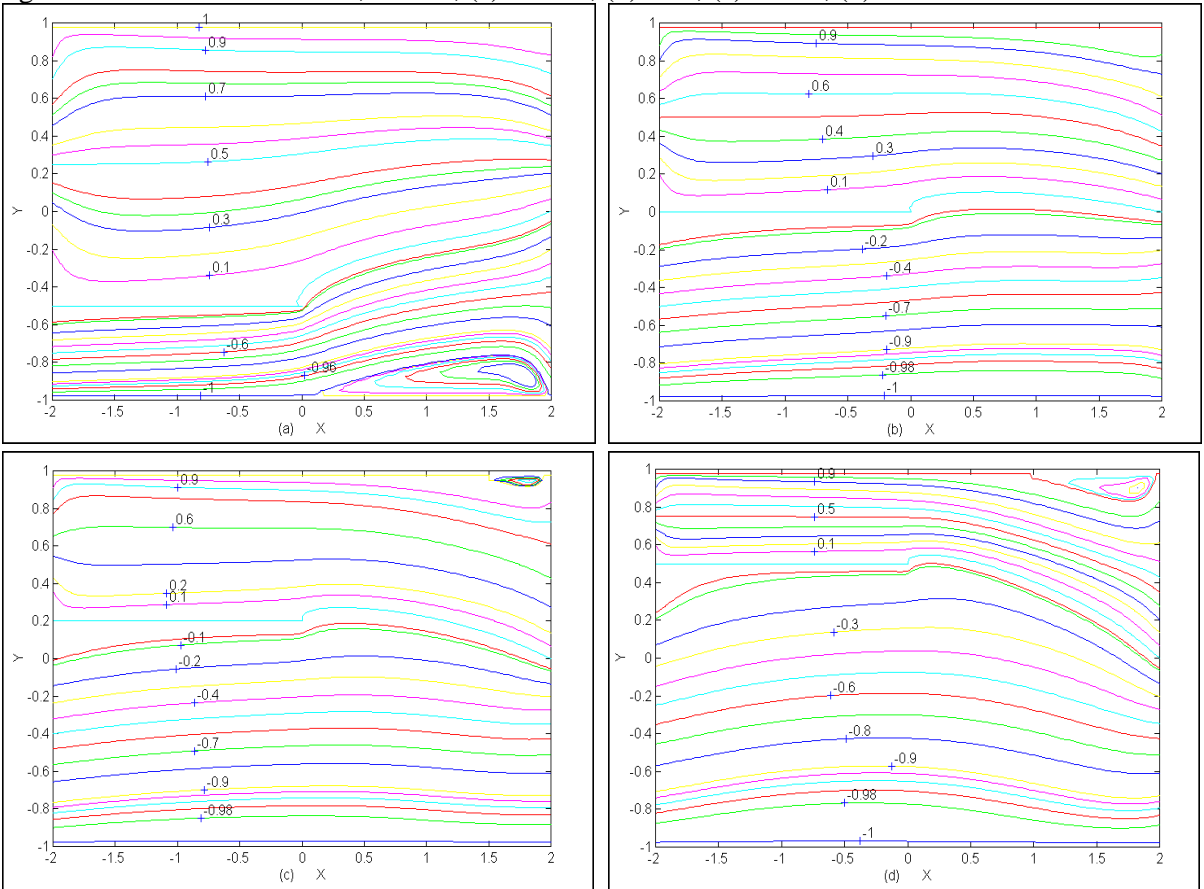

Figure 5. Streamlines for $L'=4$, R=1000, (a) $\alpha=-0.5$, (b) $\alpha=0$, (c) $\alpha=0.2$, (d) $\alpha=0.5$.

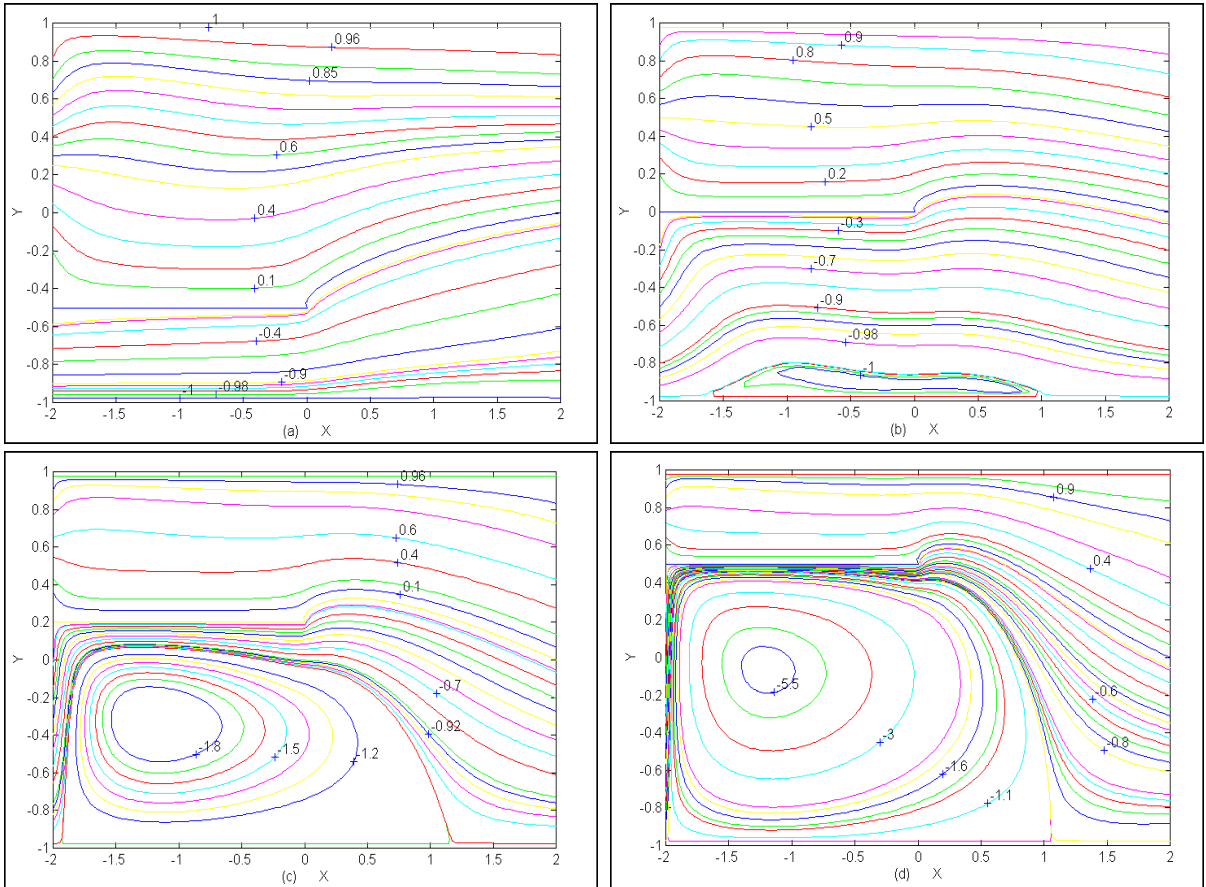

Figure 6. Streamlines for $L'=4$, $R=10^5$, (a) $\alpha=-0.5$, (b) $\alpha=0$, (c) $\alpha=0.2$, (d) $\alpha=0.5$.

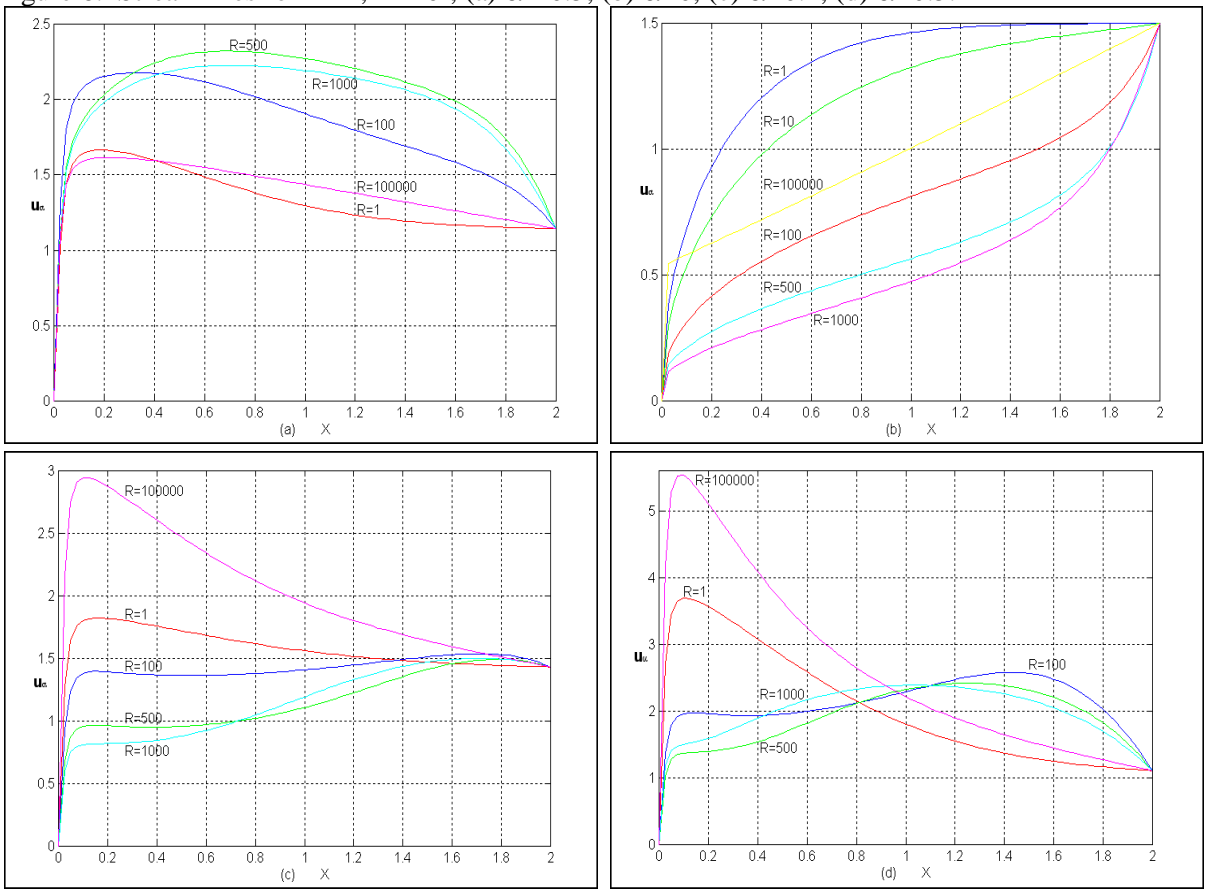

Figure 7. The variation of $u_{cl}$ for various values of R and (a) $\alpha=-0.5$, (b) $\alpha=0$, (c) $\alpha=0.2$, (d) $\alpha=0.5$.

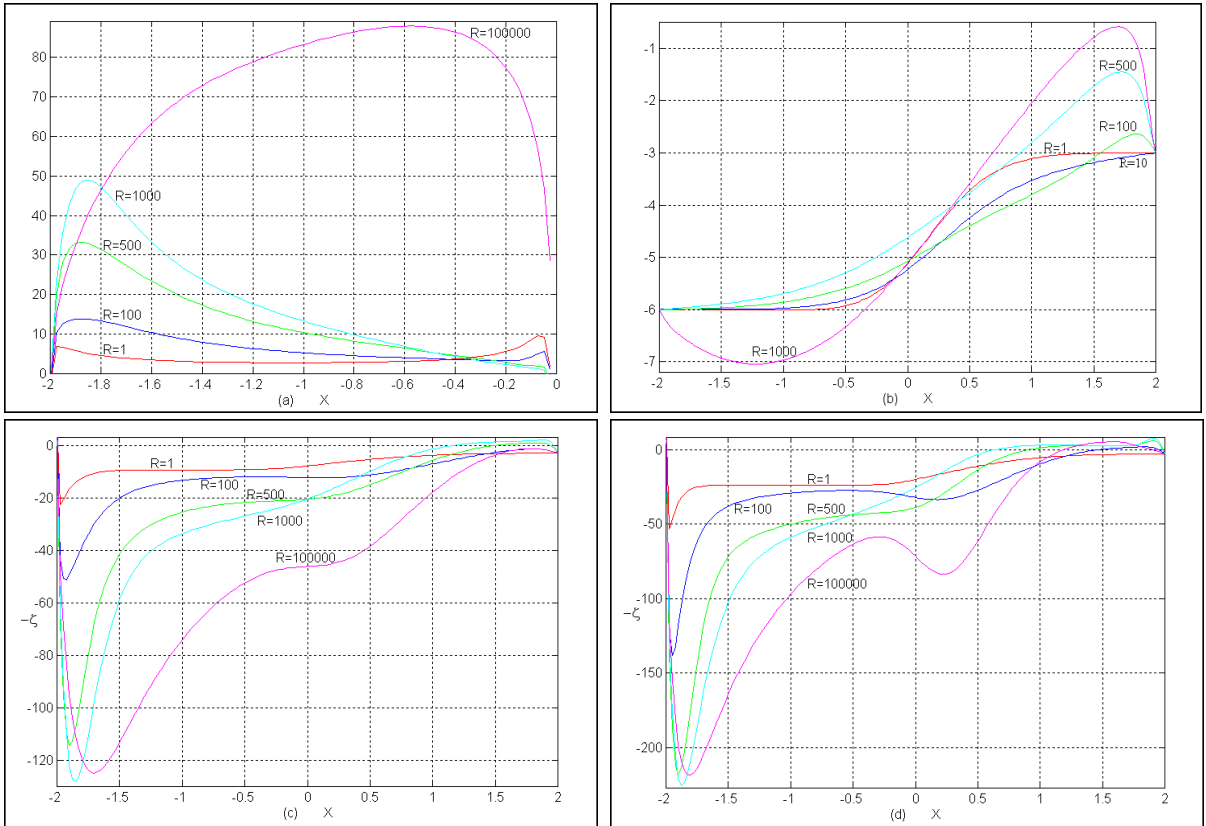

Figure 8. The variation of wall vorticity on upper plate of channel for various values of R and
(a) α=-0.5, (b) α=0, (c) α=0.2, (d) α=0.5.

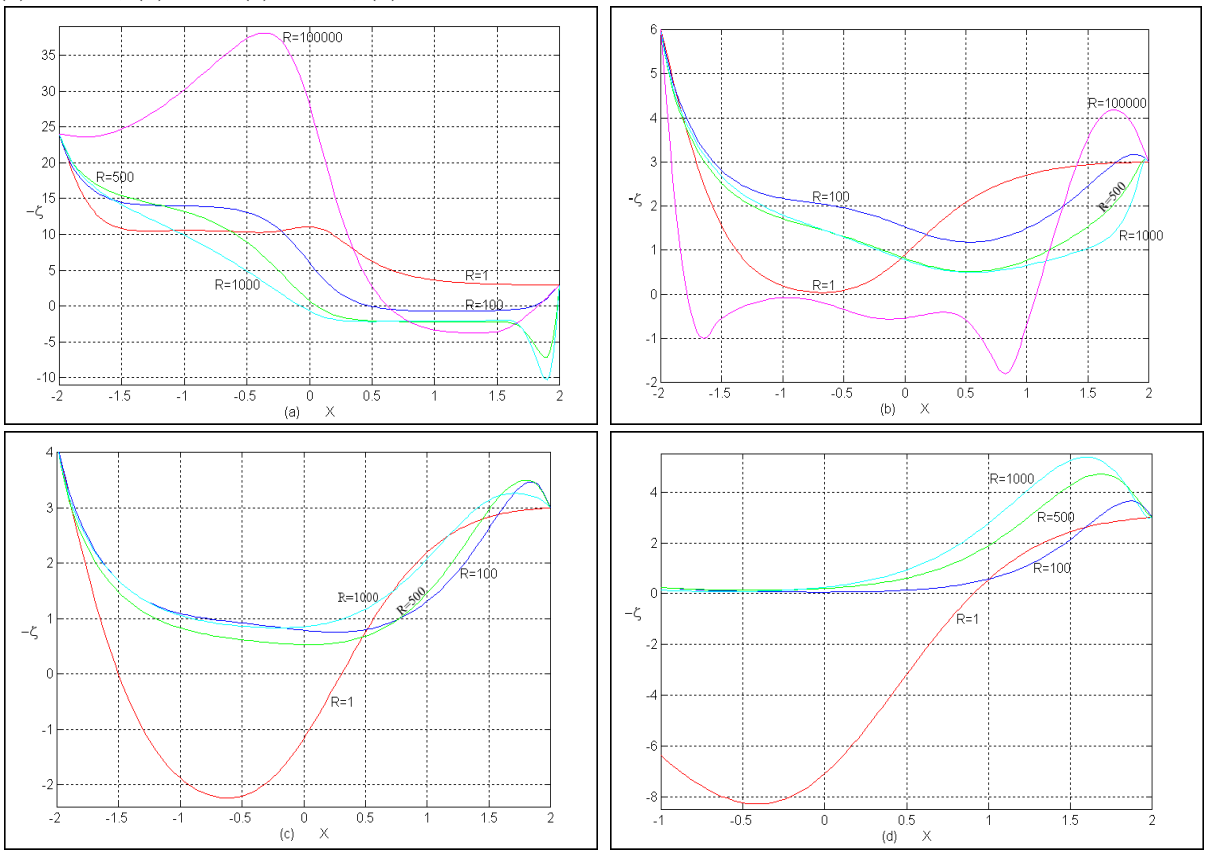

Figure 9. The variation of wall vorticity on lower plate of channel for various values of R and
(a) α=-0.5, (b) α=0, (c) α=0.2, (d) α=0.5.

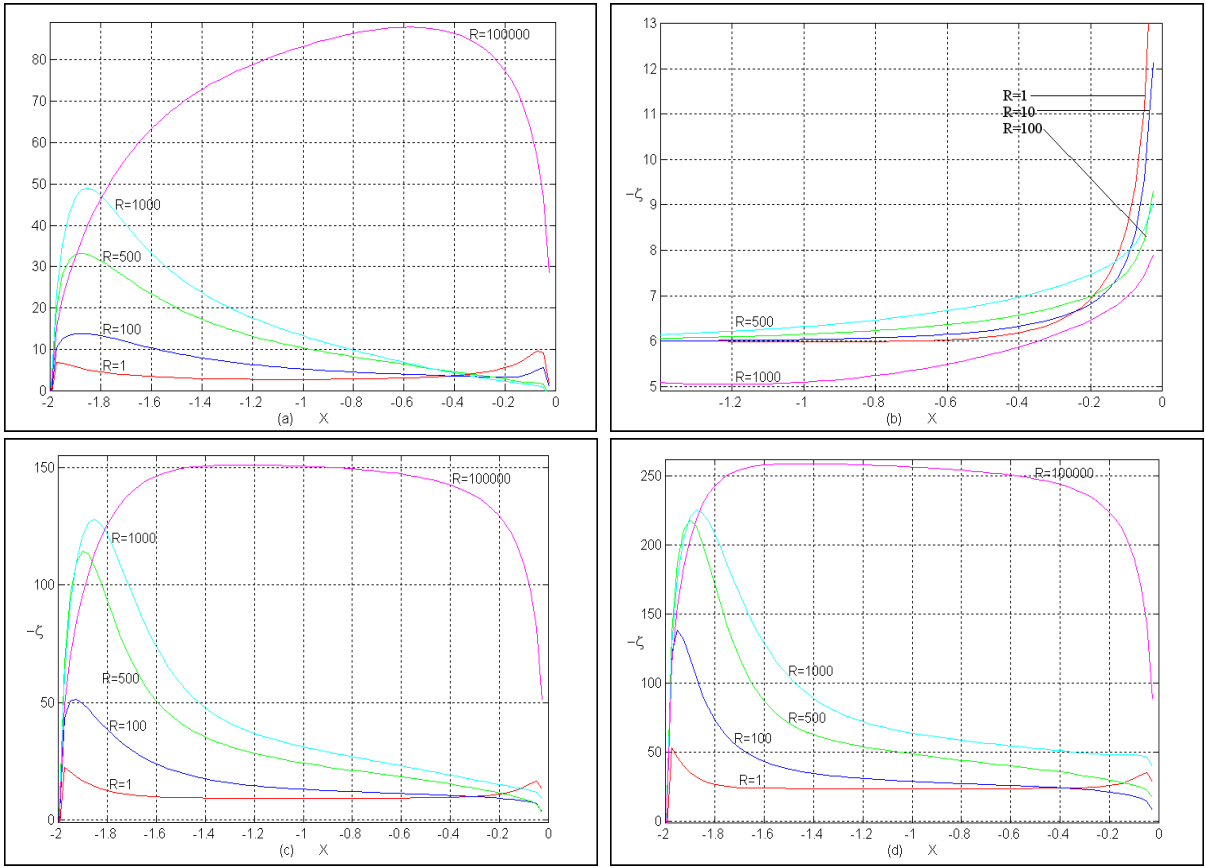

Figure 10. The variation of wall vorticity on splitter of channel for various values of R and (a) α=-0.5, (b) α=0, (c) α=0.2, (d) α=0.5.

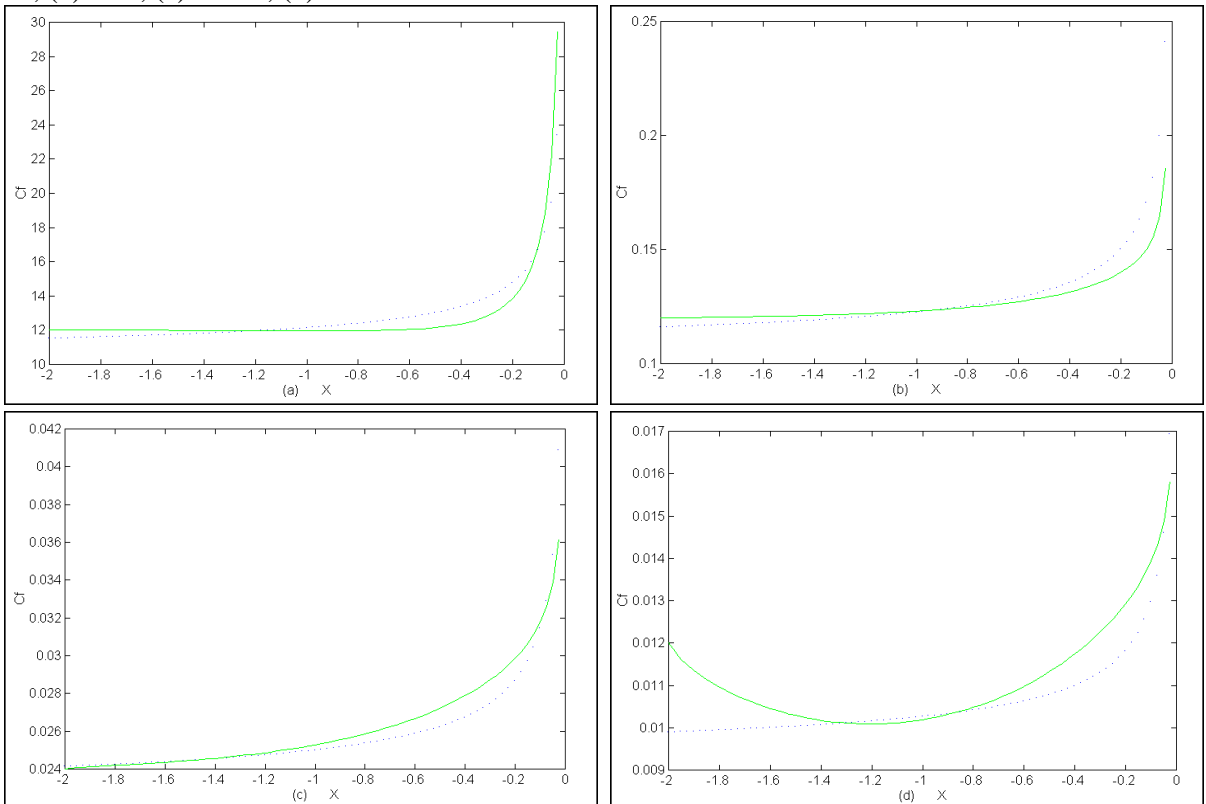

Figure 11. Comparison of present numerical solution for shear stress on the splitter for symmetric case with analytic solution of Badr [1] (shown here dotted) when (a) R=1, (b) R=100, (c) R=500, (d) R=1000.

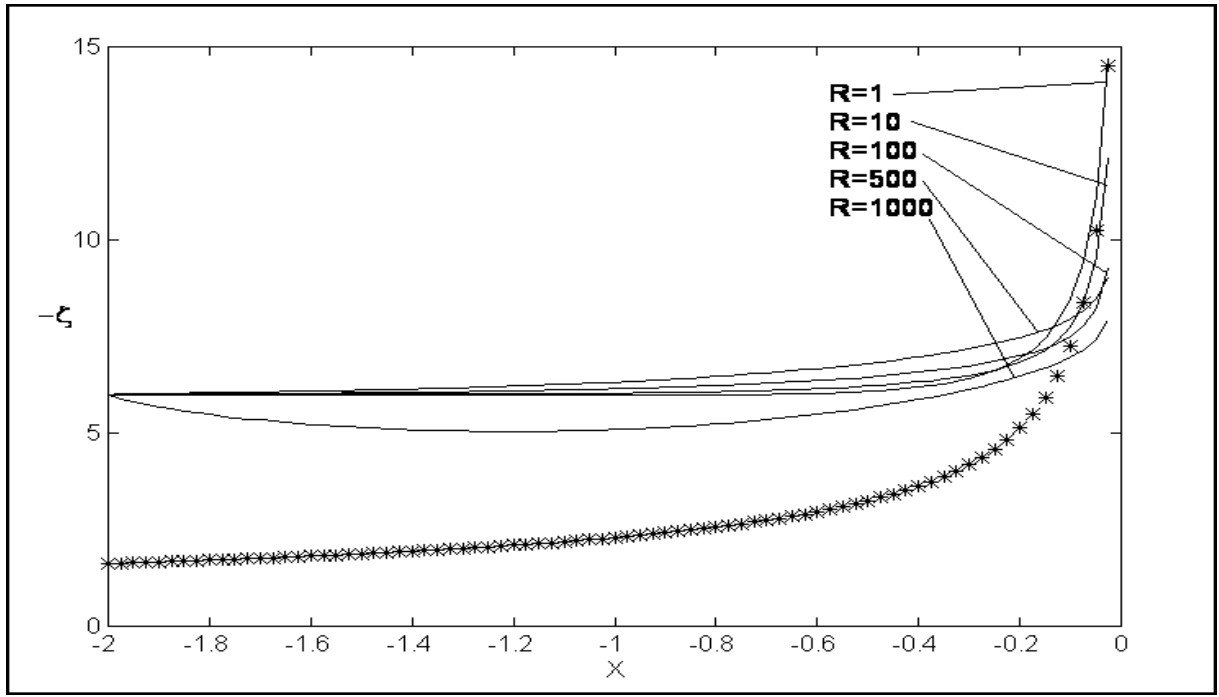

Figure 12. Comparison of present results with analytical results of Badr [1] (here shown as *) for the variation of wall vorticity on splitter of channel for various values of R and (a) α=-0.5, (b) α=0, (c) α=0.2, (d) α=0.5

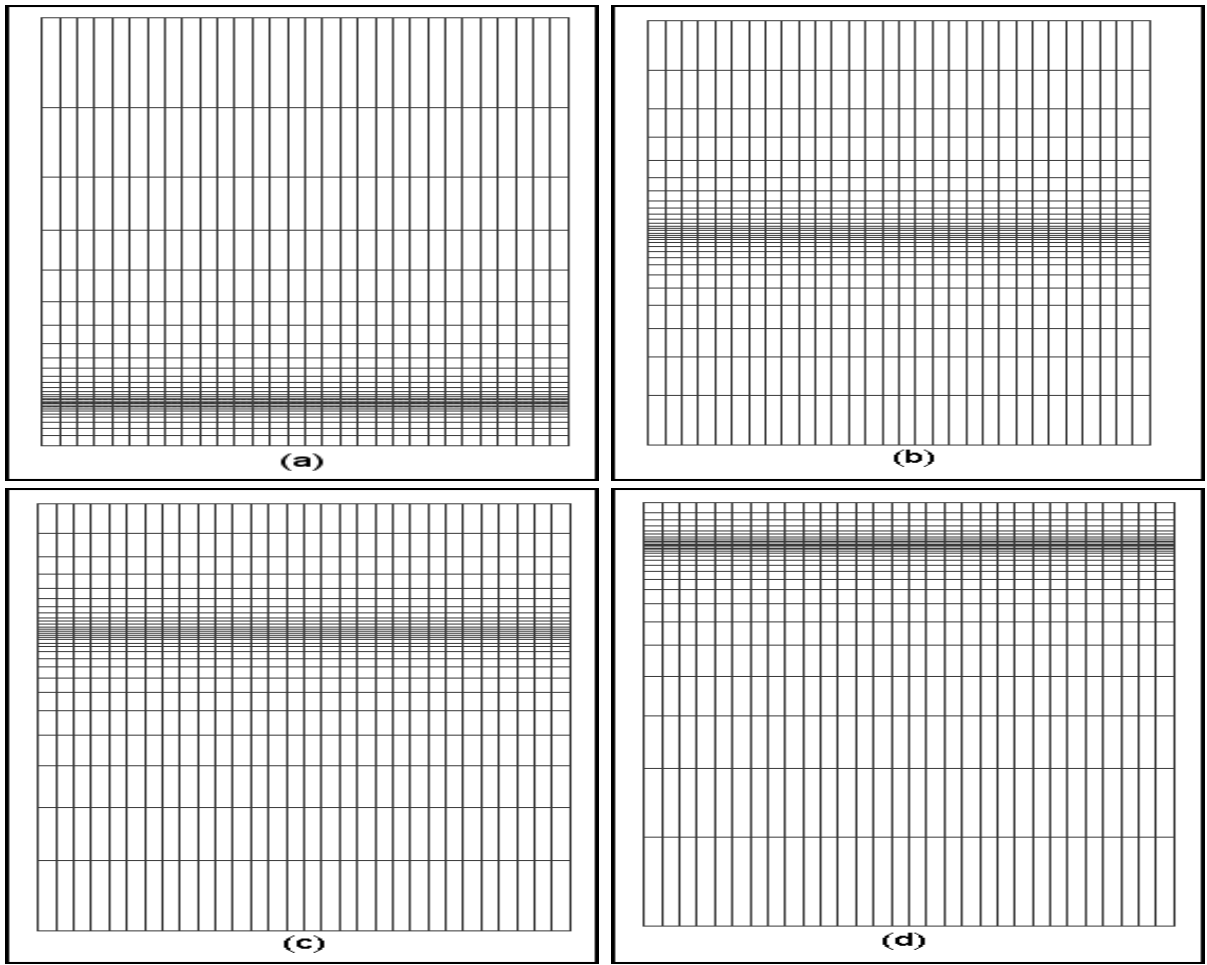

Figure 13. Clustering in the vicinity of splitter plate for different values of α considered.